\documentclass[12pt,a4paper]{article}  
\newtheorem{lemma}{Lemma}

\newcommand{\M}{{\cal M}}
\newcommand{\N}{{\cal N}} 
\newcommand{\NN}{\widehat{\cal N}}
\newcommand\bm[1]{\mbox{\boldmath$#1$}} 
\newcommand{\oo}{(\pm)}
\newcommand{\op}{(+)} 
\newcommand{\om}{(-)}
\author{Marc Mars\thanks{Supported by the European Union under a
Marie Curie Fellowship.}
\thanks{e-mail: marc@galileo.thp.univie.ac.at}~ and Walter Simon
\thanks{Supported by Jubil\"aumsfonds der \"Osterreichischen Nationalbank, 
project 6265.} \thanks{e-mail: simon@galileo.thp.univie.ac.at}\\
Institut f\"ur Theoretische
Physik der Univ. Wien,\\ Boltzmanngasse 5,\\
 A-1090 Wien, Austria.}

\title{A Proof of Uniqueness of the Taub-bolt Instanton}
\date{}
\begin{document}

\maketitle

\begin{abstract} 
We show that the Riemannian Schwarzschild and the ``Taub-bolt''
instanton solutions are the only spaces $({\cal M}, g_{\mu\nu})$
such that 
\begin{itemize}
\item ${\cal M}$ is a 4-dimensional, simply connected manifold
with a Riemannian, Ricci-flat $C^2-$metric
$g_{\mu\nu}$ which admits (at least) a 1-parameter group
$\mu_{\tau}$ of isometries without isolated fixed points on $\M$.
\item The quotient $(\M \setminus {\cal L}_{\M})/\mu_{\tau}$ (where 
${\cal L}_{\M}$ is the set of fixed points of $\mu_{\tau}$)
is an asymptotically flat manifold,
and the length of the Killing field corresponding to
$\mu_{\tau}$ tends to a constant at infinity.
\end{itemize}
\vfill
MSC number 53C25\\
PACS numbers 0420, 0240

\end{abstract}

\newpage

\section{Introduction}

Attempts of estimating the path integral of Quantum Gravity via
the stationary phase approximation motivated the study of
``instantons'', i.e.  Riemannian, Ricci-flat metrics which are
regular everywhere.  Gibbons and Hawking \cite{GH} distinguished instantons
having at least a one-parameter group of isometries according to
whether the isometry has isolated fixed points (``nuts'') or
two-dimensional subsets of fixed points (``bolts''). The mathematical
analogy between instantons with  fixed points and Lorentzian
solutions with Killing horizons raises the
question whether the known uniqueness proofs for stationary 
black hole solutions
can be carried over to the Riemannian case.  In fact, 
for asymptotically flat solutions with up to two nuts and no
bolts, a uniqueness result for the Riemannian Kerr metric has been obtained
\cite{WS} by adapting and generalizing Israel's proof of uniqueness of the
Schwarzschild solution \cite{WI}.

The present work concerns the problem of finding uniqueness
results for the 1-parameter family of Schwarzschild instantons and for 
the 1-parameter family of so-called Taub-bolt instantons (found by Page \cite{DP} by 
``Euclideanizing'' the Taub-NUT spacetime \cite{TN}) which read as follows,
respectively,
\begin{eqnarray}
\label{Schwarz}
ds^2 & = & \frac{r-m}{r+m} d\tau^2 + \frac{r+m}{r-m}
\left(dr^2 + \left (r^{2} - m^{2} \right)d\Omega^2 \right) 
\\ \nonumber \\
\label{Taubbolt}
ds^2 & = & \frac{\left (r - 2|n| \right)\left(r - \frac{1}{2}|n| \right)}
{(r^2 - n^2)} \left (d\tau + 2 n\cos \theta d \phi \right )^2 + {}
\nonumber\\
& & {} + (r^2 - n^2)\left (\frac{dr^2}{\left(r - 2|n| \right) 
\left(r - \frac{1}{2}|n| \right)} +  d\Omega^{2} \right),
\end{eqnarray}
where $m\ge 0$ and $n \ne 0$ are constants and $d\Omega^{2}$ is the round metric
on the 2-sphere. In (\ref{Schwarz}) the subcase $m=0$ is just
the Euclidean metric on $R^4$ whereas for $m > 0$ the coordinate
$\tau$ is periodic with range $0 \le \tau < 8\pi m$, and
$r \ge m$. Regarding (\ref{Taubbolt}), $\tau$ is periodic
with range $0 \le \tau < 8\pi n$, and $r \ge 2|n|$.
These spaces are probably unique in the following sense. \\ \\
{\bf Conjecture}~{\it The Riemannian Schwarzschild and the 
``Taub-bolt'' instanton solutions are the only spaces 
$(\M, g_{\mu\nu})$ such that 
\begin{description}
\item[C1.]
$\M$ is a 4-dimensional manifold
with a Riemannian, Ricci-flat $C^{2}-$metric $g_{\mu\nu}$ 
which admits (at least) a 1-parameter group $\mu_{\tau}$ of isometries. 
\item[C2.] $\mu_{\tau}$ has no isolated fixed points on $\M$. 
\item[C3.] $\M$ is simply connected.
\item[C4.] $(\M,g_{\mu\nu})$ is asymptotically flat (AF) or
asymptotically locally flat (ALF), and the
trajectories of $\mu_{\tau}$ have bounded length at infinity.
\end{description} }
In defining {\it AF} and {\it ALF} we may follow \cite{GPRP}. 
(Note that according to this definition, flat space is not {\it AF}
but asymptotically Euclidean).

While it would be desirable to prove this conjecture, we have obtained a
closely related uniqueness result.
To formulate the latter we introduce the set 
 $\N = \pi (\M \setminus {\cal L}_{\M} )$ of non-trivial orbits of
$\mu_{\tau}$, where ${\cal L}_{\M}$ denotes the set of fixed points of
$\mu_{\tau}$ and 
$\pi: \M \rightarrow \M/ \mu_{\tau}$ is the canonical projection.
In general, the space $\N$ of
Killing orbits is what has been called a V-manifold \cite{IS}, a
Satake-manifold \cite{PM} or an orbifold \cite{BGM}. However, in some
cases, (such as for Schwarzschild and Taub-bolt), $\N$ actually has
the structure of a (standard) manifold. In this case
there exists a one-to-one
correspondence between tensors on $\N$ and tensors on
$\M \setminus {\cal L}_{\M}$ which are orthogonal with respect to every index 
to the Killing field $\xi^{\mu}$ corresponding to
$\mu_{\tau}$, and have vanishing Lie derivative along $\xi^{\mu}$.
In particular,  the symmetric covariant tensor
$g_{\mu\nu} - V^{-2}\xi_{\mu}\xi_{\nu} $ 
on $\M$ (where $V = (g_{\mu\nu}\xi^{\mu}\xi^{\nu})^{1/2}$ is the norm of
$\xi^{\mu}$) can be ``pulled back'' to give a metric on $\N$ which we
call $g_{ij}$. With this metric $(\N,g_{ij})$ is a Riemannian manifold.

Our theorem reads as follows.\\ \\
{\bf Theorem}~{\it The Riemannian Schwarzschild and the
Taub-bolt instanton solutions are 
the only spaces $({\cal M}, g_{\mu\nu})$ such that 
\begin{description}
\item[T1.] ${\cal M}$ is a 4-dimensional manifold
with a Riemannian, Ricci-flat $ C^{2}-$metric $g_{\mu\nu}$ 
which admits (at least) a 1-parameter group $\mu_{\tau}$ of
isometries. Moreover, the set $\N$ of non-trivial orbits of
$\mu_{\tau}$ is a manifold.
\item[T2.] $\mu_{\tau}$ has no isolated fixed points on $\M$.
\item[T3.] $\M$ is simply connected.
\item[T4.] $(\N,g_{ij})$ is asymptotically flat (AF) 
and the norm of the Killing field corresponding
to $\mu_{\tau}$ tends to a constant at infinity.
\end{description} }

Note that conditions {\bf C2}-{\bf T2} and
{\bf C3}-{\bf T3} agree
whereas the other conditions of the theorem have
similar but different counterparts in the conjecture.
In particular, in {\bf T1} we have included the requirement 
that $\N$ is a manifold. Actually, this latter requirement is 
probably spurious due to the strong condition {\bf T2}.
 We will come back to this point in Sect. 5.
As to {\bf T4}, the notion of asymptotic flatness for $(\N,g_{ij})$
will be defined in Sect 3. We remark that condition ${\bf T4}$ allows 
$(\M,g_{\mu\nu})$ to be flat Euclidean space, in contrast to ${\bf C4}$.

Our requirement on the metric to be $C^2$ permits the
introduction of harmonic coordinates, in terms of which the
condition of Ricci flatness is an elliptic equation for the
metric, and so the latter is analytic. An elementary result 
(Sect. VI, Proposition 1.1. of \cite{KN1}) shows that in 
this case the Killing field $\xi^{\mu}$ is $C^{\infty}$. But
since $\xi^{\mu}$ also satisfies an elliptic equation in Ricci flat
spaces, it is analytic as well.
It is then also possible to find an analytic
atlas on ${\cal M}$ of the form $(t, x^{i})$ where $t$ is a
function of the group parameter $\tau$, and $x^{i}$ are coordinates
on $\N$ \cite{HM}. It will turn out to be convenient to use this atlas
in particular in the asymptotic analysis (Sect. 3).

Our proof follows the method of Bunting and Masood-ul-Alam
for proving uniqueness of the Schwarzschild solution among
the static, asymptotically flat vacuum black holes \cite{BM}
with bifurcate horizons.
They construct a complete space $\overline{\N}$ by gluing together two
copies of a $t=const.$ slice along the bifurcation surface of the
horizons, and by performing a 1-point
compactification of one of the ends.
By a suitable conformal transformation, $\overline{\N}$ can be
endowed with a metric of vanishing mass, which is
sufficiently regular such that the limiting case of the positive
mass theorem \cite{SY} can be applied. This yields that $\overline{\N}$
is flat and the original space is Schwarzschild.

In our case the orbit space $\N$ takes the role of the
$t=const.$ slices and the bolts take the role of the bifurcation surfaces.
To follow the strategy of \cite{BM} we first (in Sect. 2) have to
study carefully the local geometry of $\N$ in the
neighbourhood of a bolt. We then (in Sect. 3) perform a detailed analysis of
the asymptotic properties of $\N$. Sect. 4 contains a preliminary
result on the global geometry and the proof of the theorem.
In the final section we discuss the options of proving the conjecture stated
above, and of improving or generalizing our theorem by relaxing conditions {\bf T1} 
and {\bf T2}.

\section{Properties of Bolts}

The set ${\cal L}_{\cal M}$ of fixed points of $\mu_{\tau}$ has
the following structure (c.f. Sect. 2 of \cite{GH}). If 
$q \in {\cal L}_{\cal M}$ is 
isolated it is called a ``nut'' (after the Euclidean Taub-NUT
metric \cite{SH}). We exclude nuts by assumption ${\bf T2}$ of
our theorem. The remaining possibility is that each connected
component of ${\cal L}_{\cal M}$ is a two-surface called ``bolt''.
At every $q \in {\cal L}_{\cal M}$  the differential
$\mu_{\tau \ast}$ acts as the identity on a 2-dimensional subspace
$T_{q}^{-}{\cal M}$ of the tangent space $T_{q}{\cal M}$ and as a
rotation on the orthogonal subspace $T_{q}^{+}{\cal M}$, with period 
$\tau_{+} = 2\pi/\kappa$. $\kappa$ (called the ``gravity'' of
the bolt) is constant on each bolt and also appears in the
matrix representation of $\nabla_{\mu}\xi_{\nu}$ in an
orthonormal frame (c.f. Theorem 5.3 of \cite{SK}).

In a normal neighbourhood $U_q$ of a point $q$ of a bolt, we
choose normal coordinates $\{z^{\alpha}\}$
as follows. Choose a basis
of orthonormal vectors $\{\vec{v}_{\alpha} \}$
in $T_q \M$ such that $\{\vec{v}_0,\vec{v}_1\}$ span $T_q^{+}{\cal M}$ 
and  $\{\vec{v}_2,\vec{v}_3 \}$ span $T_q^{-}{\cal M}$. The normal
coordinates $\{z^{\alpha}\}$ of a point $q$ are
$\exp(z^{\alpha} \vec{v}_{\alpha})=q$.
The commutativity of $\mu_{\tau}$
and the exponential map, viz. $\exp (\mu_{\tau \ast} X) = \mu_{\tau}(\exp X)$
for $X \in T_{q} \M$ implies that the action of $\mu_{\tau}$ is linear in these
coordinates, i.e. $\mu_{\tau}^{\alpha} (z) =  
\left (\mu_{\tau \ast}\right )^{\alpha}_{\beta} (q) z^{\beta}$.
For the Killing vector we get
\begin{equation}
\xi^{\alpha} (z) = 
\left . \frac{d}{d\tau} \mu_{\tau}^{\alpha} (z) \right
|_{\tau=0} = 
\kappa \left (z^1 \partial_{z^0} - z^0 \partial_{z^1} \right ). 
\label{Kill}
\end{equation}
We now obtain the form of the metric near a bolt in normal coordinates.

\begin{lemma}
\label{metric} 
Let $q$ be an arbitrary point on a bolt and $U_{q}$ a 
normal neighbourhood of $q$. 
Consider an arbitrary $C^{2}-$metric $g_{\alpha\beta}$ defined on $U_q$
which is invariant under the action of $\mu_{\tau}$.
Then, in the normal coordinate system introduced above, the
metric takes the form
(in matrix notation and with $\dagger$ denoting transposition)
\begin{eqnarray}
\label{matrix}
g(z)  = 
\left ( \begin{array}{ll}
            S^{\dagger}(z)   &  0 \\
            0  &  {\bm I}_2
            \end{array}
\right ) \left .
\left ( \begin{array}{ll}
            A   &   B \\
            B^{\dagger}  & C
            \end{array}
\right ) \right |_{\rho(z),z^2,z^3}
\left ( \begin{array}{ll}
            S(z)  &  0 \\
            0 &  {\bm I_2}
            \end{array}
\right ),
\end{eqnarray}
where $\bm{I_2}$ is the $2 \times 2$ unit matrix,
$S$ is the rotation 
\begin{eqnarray}
\label{Smat}
S(z) = \frac{1}{\rho(z)} \left ( \begin{array}{rr} z^{0} & z^{1} \\ 
-z^{1} & z^{0} \end{array} \right ),
\end{eqnarray}
$\rho(z) \equiv  + \sqrt{(z^0)^2 + (z^1)^2}$, and the $2\times2$~ matrices
$A, B, C$ are $C^2$ functions of three variables $\rho,~z^2~and~z^3$
in a domain $\rho \geq 0$. Furthermore,
\begin{eqnarray}
A(\rho,z^2,z^3)  = 
 a\left(z^2,z^3 \right)\bm{I_2} +  O\left(\rho^2 \right ), \hspace{2mm}
B(\rho,z^2,z^3)  =  \rho B^{(1)}\left(z^{2}, z^{3} \right) +
O\left (\rho^2 \right ), \nonumber \\
C(\rho,z^2,z^3)  =   C^{(0)}\left (z^2,z^3 \right) + O \left (\rho^2 \right ),
\hspace{3cm} \label{expansion}
\end{eqnarray}
where $a\left(z^2,z^3 \right)$ is a function with $a(0,0) = 1$, and  
$C^{(0)}(0,0) = \bm{I_2}$.
\label{Taylor}
\end{lemma}
{\it Proof.} 
The action of the isometry group in the normal coordinates
$\{ z^{\alpha} \}$ is
$ \mu_{\tau} (z)^{\alpha} = {H_{\tau}}^{\alpha}_{\beta}
z^{\beta}$, where $H_{\tau}$ is (in matrix notation)
\begin{eqnarray*} 
 H_{\tau}  = \left ( \begin{array}{cc}
        R_{\tau} & 0 \\
        0 & \bm{I_2} 
        \end{array}
\right ),  \hspace{1cm} R_{\tau} = \left (
\begin{array}{cc}
\cos (\kappa \tau) & \sin (\kappa \tau) \\
-\sin (\kappa \tau) & \cos (\kappa \tau) 
\end{array}
\right ).
\end{eqnarray*}
Invariance of the metric under $\mu_{\tau}$ means
$g_{\alpha\beta} (z) = 
 \left (\mu_{\tau}^{\mu} (z) \right )_{,\alpha}
\left (\mu_{\tau}^{\nu} (z) \right )_{,\beta}
g_{\mu\nu} \left (\mu_{\tau} (z) \right )$, where comma denotes
partial derivative. Hence
$g(H_{\tau} z) = H_{\tau} g(z) H_{\tau}^{\dagger}$, where
$(H_{\tau} z)^{\alpha} = {H_{\tau}}^{\alpha}_{\beta} z^{\beta}$.
The matrix $S(z)$ is defined only at points with $\rho(z) \neq 0$ (although
the expression (\ref{matrix}) in the lemma makes sense also at $\rho(z) = 0$
because of the expansion (\ref{expansion})). Let us define a matrix
$\tilde{g}(z)$ by
\begin{eqnarray*}
\tilde{g}(z) = \left ( \begin{array}{cc}
                S(z) & 0 \\
                0 & {\bm I_2} 
                \end{array}
\right )
g (z) 
\left ( \begin{array}{cc}
                S^{\dagger}(z)  & 0 \\
                0 & {\bm I_2}
        \end{array}   
\right ) \hspace{1cm} \mbox{for } z \mbox{ such that } \rho(z) \neq 0. 
\end{eqnarray*}
Using the identity  $S (H_{\tau} z ) = S ( z ) R^{-1} (\tau)$, the invariance
equation becomes simply $\tilde{g} (H_{\tau} z ) = \tilde{g} (z)$ at
points where $\rho(z) \neq 0$. This equation motivates the 
definition of $2 \times 2$ matrices $A,B$ and $C$, which are functions of three
variables $\rho,z^2,z^3$ on a domain where $\rho \geq 0$,
through the equation
\begin{eqnarray}
\left . \left (\begin{array}{cc}
                               A & B \\
                               B^{\dagger} &  C 
                               \end{array}
\right ) \right |_{\rho,z^2,z^3} = 
g(z) |_{z^0 = \rho, z^1=0, z^2,z^3}, \hspace{1cm}
\mbox{where} \hspace{3mm} \rho  \geq 0. 
\label{defi}
\end{eqnarray}
Since the right-hand side is a $C^2$ function of its arguments,
so are $A,B$ and $C$. Using the form of $S$ and the invariance
equation, we obtain for any $\rho < 0$
\begin{eqnarray}
g(z) |_{z^0= \rho, z^1=0,z^2,z^3} =                                 
 \left . \left (\begin{array}{cc}
                               A & -B \\
                               -B^{\dagger} &  C 
                               \end{array}
\right ) \right |_{- \rho,z^2,z^3}.
\label{nega}
\end{eqnarray}
This equation together with the continuity of $g(z)$ at $\rho(z)=0$
implies that $B(0,z^2,z^3)=0$. Let now
$a,b =0,1$ and $A,B=2,3$. Using again the form of $S(z)$ and the invariance
equation, we have, for $\rho >0$,
\begin{eqnarray*}
g_{ab}(z) |_{z^0=0, z^1=\rho, z^2,z^3} = \left . \left (
\begin{array}{cc}
A_{11}  & -  A_{01}  \\
-A_{01}  &   A_{00}  
\end{array} \right ) \right |_{\rho,z^2,z^3}.
\end{eqnarray*}
The limit $\rho \rightarrow 0$ in this expression makes sense, and using 
definition (\ref{defi}) we
obtain $A(0,z^2,z^3) = a(z^2,z^3) \bm{I_2}$
where $a(z^2,z^3)$ is a $C^2$
function. Taking partial derivatives of  (\ref{defi}) and (\ref{nega})
with respect to $\rho$, performing the limit $\rho\rightarrow 0$ and
using that $g(z)$ has continuous derivatives, 
we easily get 
$\left .\partial_{\rho} A \right |_{\rho=0}=0$ and 
$\left . \partial_{\rho} C \right  |_{\rho=0}=0$.
Noticing that $g_{\mu\nu} (z=0) = \delta_{\mu\nu}$, the 
lemma follows by Taylor expansion.\\

We denote the set of all orbits of $\mu_{\tau}$
(including its fixed points ${\cal L_{\cal M}}$) by $\NN$, and  
define $\pi({\cal L}_{\M}) = {\cal L}_{\NN}$. The latter set has
the following structure.

\begin{lemma}
Let $(\M, g_{\mu\nu})$ satisfy conditions {\bf T1} and {\bf T2}
of the theorem.

Then ${\cal L}_{\NN}$ is a smooth, two-dimensional boundary of $\N$. 
Moreover, $g_{ij}$ has a $C^{2}$-extension to ${\cal L}_{\NN}$ and
${\cal L}_{\NN}$  is totally geodesic in $(\NN, g_{ij})$.
\label{bound}
\end{lemma}

{\it Proof.} In a normal neighbourhood of a point $q$ of a bolt we use
normal coordinates $\{z^{\alpha} \}$ as introduced above Lemma \ref{Taylor}.
Let $a,b=0,1$,\, $A,B=2,3$, and define $y^0= z^2$, $y^1 = z^3$.  
Using the form of the metric in Lemma \ref{Taylor} we
easily obtain, at points where $\rho(z) \neq 0$,
\begin{eqnarray*}
\xi^{a} \xi^{b} g_{ab}= \kappa^2 \rho^2 A_{11}, \hspace{1cm}
g_{ab} \xi^a dz^b  = -\kappa \rho A_{1b} \bm{\alpha^b}, \hspace{1cm}
g_{a B} \xi^a dz^B  = - \kappa \rho B_{1b} dy^b, \\
g_{ab} dz^a dz^{b} = A_{ab} \bm{\alpha}^{a} \bm{\alpha}^b,
\hspace{1cm} 
g_{aB} dz^a dz^{B} = B_{ab} \bm{\alpha}^{a} dy^b,
\hspace{2cm}
\end{eqnarray*}
where we have introduced two one-forms $\bm{\alpha}^0= d \rho =
(z^0 dz^0 + z^1 dz^1)/\rho$ and $\bm{\alpha}^1 = 
(-z^1 dz^0 + z^0 dz^1)/\rho$. 
It is now straightforward to show that the symmetric tensor
$ds^2_g= (g_{\mu\nu} - V^{-2}\xi_{\mu}\xi_{\nu})dz^{\mu}dz^{\nu}$
takes the form
\begin{equation}
\label{proj}
ds_g^2=
\frac{\det A}{A_{11}} d\rho^2 + 2 \left ( B_{0a} -
\frac{A_{01}} {A_{11}} B_{1a} \right ) d\rho \, dy^a 
+ \left (C_{ab} - \frac{B_{1a} B_{1b}}{A_{11}} \right ) dy^a dy^b.
\end{equation}
Since $\rho$ and $y^a$ are constant along the Killing orbits, they are
suitable coordinates
on $\NN$ ($\rho$ is of course restricted to be non-negative) and
therefore the metric $g_{ij}$ on $\N$ is given exactly by
expression (\ref{proj}), where now $d\rho$ is no longer 
$\bm{\alpha^0}$ but the differential of a coordinate.
The boundary of $\NN$ which is given locally by $\rho=0$  is two-dimensional
and  coincides with ${\cal L}_{\NN}$ by construction.
Lemma \ref{Taylor} implies $ds^2_{g}  |_{\rho=0} =
a (y^c) d\rho^2 + C^{(0)}_{ab}(y^c) dy^a dy^b$. Since
$A, B$ and $C$ are $C^2$ functions of $\rho,z^2$ and $z^3$,
it follows that
the metric $g_{ij}$ extends at least in
a $C^2$ manner to $\rho=0$. A trivial calculation shows
that the second fundamental form of  
${\cal L}_{\NN}$ in $(\NN, g_{ij})$ vanishes, which is 
equivalent to ${\cal L}_{\NN}$ being totally geodesic.
Finally, the smoothness
of $\NN$ follows from the smoothness of the geodesics.

\section{Asymptotic structure}
Here we first define asymptotic flatness of
the manifold $(\N,g_{ij})$ (contained in assumption {\bf T4} of the theorem).
We then use assumption {\bf T1}, in particular Ricci flatness,
together with asymptotic properties, to introduce the ``twist''
(-scalar) of the Killing field on an ``end''  $\N^{\infty}$.
Next, we adapt from \cite{NM} two technical lemmas 
(Lemmas \ref{ellinv} and \ref{Lapinv} below) on the
inversion of a certain elliptic operator and of the flat Laplacian.
These results will be used to establish falloff properties of the
metric on $\N$ and of quantities constructed from the norm and
the twist of $\xi^{\mu}$ (Lemma \ref{as}), and to show the existence of a 
compactification of the end of $\N$ (Lemma \ref{comp}).
Most parts of the proofs of these lemmas parallel closely 
the treatment of the Lorentzian case (\cite{BC}, \cite{BS}, \cite{SB}).\\ \\
{\bf Definition} 
{\it The manifold $(\N, g_{ij})$ is called  asymptotically flat iff
\begin{enumerate}
\item
The ``end'' $\N^{\infty} = \widehat \N~ \setminus$ \{a compact set\} is 
diffeomorphic to $R^{3} \setminus \, B$, where $B$ is
a ball. (The compact set is  chosen appropriately, in particular
sufficiently large).
\item 
On $\N^{\infty}$ there are coordinates such that
\begin{equation}
\label{gas}
g_{ij} - \delta_{ij} = O^{2}(r^{-\delta}) 
\qquad for~~some~~\delta > 0.
\end{equation}
\end{enumerate}
(A function $f(x^{i})$ is said to be $O^{k}(r^{\alpha})$, $k \in
N$, if $f(x^{i}) = O(r^{\alpha})$,
$\partial_{j}f(x^{i}) = O(r^{\alpha - 1})$ a.s.o. for all
derivatives up to and including the kth ones.)} \\ 

This definition might appear over-restrictive as it 
combines the topological condition $(1)$ with the purely asymptotic
condition $(2)$. 
We adopt this definition not only because it follows common
practice \cite{GPRP} but also because condition (1) is really required 
to prove the theorem in Sect. 4.
In particular, we require here that $\N$ has a single end.
Similarly, we believe that the definitions of  {\it AF} and of {\it ALF}
as given in \cite{GPRP} will be required to prove the conjecture stated
in the introduction.

On the whole of $\M$ we can introduce the twist vector
 $\omega_{\mu}= \epsilon_{\mu\nu\sigma\tau} \xi^{\nu} \nabla^{\sigma}
\xi^{\tau}$ of the Killing vector $\xi^{\mu}$.
($\epsilon_{\mu\nu\sigma\tau}$ is antisymmetric and $\epsilon_{0123} =
\sqrt{det~g})$. 
Since $\xi^{\mu}\omega_{\mu} = 0$ and since $\xi^{\mu}$ commutes
with $\omega_{\mu}$, $\omega_{\mu}$ and hence also
$\nabla_{[\mu}\omega_{\nu]}$ can be projected to a vector  
$\omega_{i}$ and to its curl
$\partial_{[i}\omega_{j]}$ on $\N$ \cite{RG}. By
Ricci-flatness, i.e. $R_{\mu\nu} = 0$, we find that $\omega_{\mu}$
and $\omega_{i}$ are curl free, i.e.
\begin{equation}
\label{Dom}
\nabla_{[\mu}\omega_{\nu]} = 0, \qquad \partial_{[i}\omega_{j]} = 0.
\end{equation}

For the rest of this section we restrict ourselves to the end
$\N^{\infty}$. Since the latter is 
simply connected, there exists a scalar $\omega$ there,
defined up to an additive constant, such that 
$\partial_{i}\omega = \omega_{i}$. 
Denoting by ${\cal D}$ the covariant derivative and by 
${\cal R}_{ij}$ the Ricci tensor with respect to $g_{ij}$,
we can decompose the condition $R_{\mu\nu} = 0$ as follows (by
changing signs appropriately in the
corresponding Lorentzian result \cite{RG})
\begin{eqnarray}
\label{gRicc}
{\cal R}_{ij} & = & V^{-1} {\cal D}_{i}{\cal D}_{j} V -
\frac{1}{2} V^{-4} (\omega_{i}\omega_{j} - 
g_{ij}g^{kl}\omega_{k}\omega_{l}),\\
\label{gLapV}
{\cal D}^{2} V & = & \frac{1}{2}V^{-3}g^{ij}\omega_{i}\omega_{j},  \\
\label{gLapom}
{\cal D}^{2} \omega & = & 3V^{-2}g^{ij}\omega_{i}{\cal D}_{j}V, 
\end{eqnarray}
and it follows from (\ref{gRicc}) and (\ref{gLapV}) that the Ricci scalar
${\cal R}$ with respect to $g_{ij}$ can be expressed as
\begin{equation}
{\cal R}  =  \frac{3}{2} V^{-4} g^{ij}\omega_{i}\omega_{j}. 
\label{trgRicc}
\end{equation}

By condition {\bf T4}, ${\cal R} = O(r^{-2-\delta})$, and
by rescaling $\xi^{\mu}$ suitably, we can achieve that
 $V \rightarrow 1$. Using (\ref{trgRicc}) we obtain 
$\omega_{i} = O(r^{-1-\epsilon})$ and so we can adjust the additive constant
in $\omega$ such that 
\begin{equation}
\label{omo1}
\omega = O^{1}(r^{-\epsilon}) \hspace{1cm} \mbox{for some positive } 
\epsilon.
\end{equation}
To analyze the full set of field equations it is convenient to
introduce some more notation.
We will employ the metric $\gamma_{ij} = V^{2} g_{ij}$ on $\N^{\infty}$
and we denote by $D_{i}$ and $R_{ij}$ the covariant derivative and the Ricci 
tensor with respect to $\gamma_{ij}$. 
We also introduce 
\begin{eqnarray}
v_{\pm} & = & \frac{1 - V^{2} \mp \omega}{1 + V^{2} \pm \omega}, \\
\Theta = 1 - v_{+}v_{-} & = & \frac{4 V^{2}}
{\left (1+ V^2 + \omega \right)\left (1+ V^2 - \omega \right)}.
\end{eqnarray}
By the asymptotic properties discussed above, the fields $v_{\pm}$ 
and $\Theta$ exist on $\N^{\infty}$, and $0 < \Theta < 1$.
Finally we define the vector field
$A_{i}=\frac{1}{2}(v_{+}D_{i}v_{-}-v_{-}D_{i}v_{+})$.
On $\N^{\infty}$, the conditions $R_{\mu\nu} = 0$ yield
\begin{eqnarray}
\label{Lapv}
D_{i} \left (  \Theta^{-1}D^{i} v_{\pm} \right ) & = & 
\pm 2\Theta^{-2}A^{i} D_{i} v_{\pm},   \\
\label{gamRicc}
R_{ij} & = & 2\Theta^{-2}(D_{(i}v_{-})(D_{j)}v_{+}).
\end{eqnarray}

We now state (modified versions of) two lemmas of N. Meyers \cite{NM}
which we require in the proof of Lemmas \ref{as} and \ref{comp}. 
Lemma \ref{ellinv} is a special case of
Corollary 1 to Theorem 1 of \cite{NM}, whereas Lemma \ref{Lapinv}
is a special case of Lemma 5 of \cite{NM}, combined with the corollary
mentioned above.

\begin{lemma}
\label{ellinv}
On an asymptotically flat end $\N^{\infty}$, let $\psi \in C^{2}$
be a solution of
\begin{equation}
g^{ij}(\partial_{i}\partial_{j} + a_{i}\partial_{j})\psi = 0
\end{equation}
with $\psi = O(r^{-\epsilon})$ \, for some real $\epsilon$
and $a_{i} = O^{1}(r^{-1-\delta})$
\, for some $\delta > 0$.

Then $\psi = O^{2}(r^{-\epsilon})$.
\end{lemma}

\begin{lemma}
\label{Lapinv}
Let $p \in N \cup \{0\}, k \in N$, $0 < \epsilon < 1$ and 
$\lambda = O^{k}(r^{-2 -p - \epsilon})$.

Then, on a domain $R^{3} \setminus B$, the equation $\triangle \phi =
\lambda$ (with $\triangle$ denoting the flat Laplacian)
has a solution $\phi_{spec} = O^{k + 2}(r^{-p-\epsilon})$.
Thus the general solution $\phi_{gen}$ which vanishes at
infinity can be written as $\phi_{gen} = \phi_{hom} + \phi_{spec}$
where $\phi_{\hom}$ is a solution of $\triangle \phi_{hom} = 0$
with terms of order $r^{-q}$, $\forall q \in N$ with $q \le p$.
\end{lemma}

We are now in the position to prove the main asymptotic result.

\begin{lemma}
\label{as}
Let $(\M, g_{\mu\nu})$ satisfy conditions {\bf T1} and {\bf T4}
of the theorem. 

Then on $\N^{\infty}$ there are coordinates $\widehat x^{i}$ 
(with $\widehat r^2 = \delta_{ij}\widehat x^{i} \widehat x^{j}$)
and constants $m_{\pm}$ and $m^{\pm}_{i}$ such that 
\begin{eqnarray}
\label{asv2}
v_{\pm} & = & \frac{m_{\pm}}{\widehat r} + 
\frac{m^{\pm}_{i}\widehat x^{i}}{\widehat r^{3}} + 
O^{2}\left (\frac{1}{\widehat r^{3}} \right), \\
\label{asgam2}
\gamma_{ij} & = & \delta_{ij} +  
\frac{m_{+}m_{-}}{\widehat r^{4}}(\widehat x^{i} \widehat x^{j} -
\widehat r^{2}\delta_{ij}) + O^{2}\left(\frac{1}{\widehat
r^{3}}\right).
\end{eqnarray}
\end{lemma}
{\it Proof.}
The first part of the proof which leads to (\ref{Vas}) is a
straightforward adaption of a result on the asymptotic behaviour
of Killing vectors obtained in the 4-dimensional framework
(\cite{BC}, Proposition 2.2).

Using (\ref{gas}), (\ref{gRicc}) and (\ref{omo1})
we get
\begin{equation}
\label{DDVas}
{\cal D}_{i}{\cal D}_{j}V = \partial_{i}\partial_{j}V -
\Gamma^{k}_{~ij}\partial_{k} V = O(r^{-2-\epsilon}).
\end{equation}
We then define $h = V^{2} + r^{2} g^{ij}{\cal D}_{i}V {\cal D}_{j}V$
and the radial derivative, $d/dr = (x^{i}/r) \partial_{i}$. 
Using (\ref{DDVas}) and Schwarz's inequality yields 
$|\frac{d}{dr} h | \le \frac{2 C h}{r}$ for some constant $C > 0$.
Upon integration, this gives $|h| \le D r^{2 C}$
for some constant $D>0$, and thus $\partial_{i}V =
O(r^{C - 1})$. 
Inserting the latter estimate in (\ref{DDVas}) gives
$\partial_{i}\partial_{j}V = O(r^{C - 2 -\epsilon})$.
Integrating again, we obtain 
\begin{equation}
\label{DV}
\partial_{i}V = E_{i} + O^{1}(r^{C - 1 - \epsilon})
\end{equation}
for some constants $E_{i}$. 
If $ C\le 1$ this can be written as $\partial_{i}V = E_{i} +
O^{1}(r^{-\epsilon})$. The latter result can also be
obtained if $C > 1$ (with some $\epsilon > 0$ possibly different
from the $\epsilon$ used above) by iterating the integration of
(\ref{DDVas}), i.e. by inserting (\ref{DV}) in
(\ref{DDVas}) once again and integrating, and by repeating this
procedure sufficiently often. One further integration then gives
$V = E_{i}x^{i} + F + O^{2}(r^{1-\epsilon})$ where $F$ is a
constant. But since $V$ is bounded by virtue of {\bf T4} and
since $\xi^{\mu}$ has been normalized so that $V$ tends to 1 at
infinity, we have $E_{i} =
0$ and $F = 1$. This now allows us to improve the iterated
integration of (\ref{DDVas}) till we end up with
\begin{equation}
\label{Vas}
V = 1 + O^{2}(r^{-\epsilon}) 
\end{equation}
where again $\epsilon > 0$ might differ from the epsilons used so far.
We now pass to the metric $\gamma_{ij} = V^{2} g_{ij}$
and use harmonic coordinates $x^{i}$ with respect to $\gamma_{ij}$
on $\N^{\infty}$. (Such coordinates exist and coincide with the
coordinates $x^{i}$ of the harmonic atlas $(t,x^{i})$ with
respect to $g_{\mu\nu}$ introduced in Sect. 1).
Applying Lemma \ref{ellinv} to (\ref{gLapom}),(\ref{omo1}) and (\ref{Vas}),
we conclude that
\begin{equation}
\label{omo2}
\omega = O^{2}(r^{-\epsilon}).
\end{equation}

Apart from irrelevant signs and numerical factors, the remaining
part of the proof is identical to the Lorentzian analysis
\cite{BS} and will only be sketched.
We write eqn.s (\ref{Lapv}) and (\ref{gamRicc}) as
\begin{equation}
\label{Lapvgam}
\triangle v_{\pm} = O(r^{-2-\epsilon}), \qquad 
\triangle \gamma_{ij} = O(r^{-2-\epsilon}).
\end{equation}
That is to say, we shift all deviations from the flat Laplacians
to the right hand side.
Applying Lemma \ref{Lapinv}, we find that there exist constants
$m_{\pm}$ such that
\begin{eqnarray}
\label{asv1}
v_{\pm} & = & \frac{m_{\pm}}{r} + 
O^{2}\left (\frac{1}{r^{1 + \epsilon}} \right ),\\ 
\label{asgam1}
\gamma_{ij} & = & \delta_{ij} + 
O^{2}\left (\frac{1}{r^{1 + \epsilon}} \right),
\end{eqnarray}
and we note that in (\ref{asgam1}) a homogeneous solution of the
Laplace equation of order $r^{-1}$ would be incompatible with the harmonic
coordinate condition.
The next step is to insert (\ref{asv1}) and (\ref{asgam1}) into 
(\ref{Lapv}) and (\ref{gamRicc}) and again to invert the
Laplacians by a trivial explicit calculation and with the help
of Lemma \ref{Lapinv}.
In general there appears a homogeneous solution in $\gamma_{ij}$
of order $r^{-2}$ which can be removed by a suitable coordinate
transformation (compatible with the harmonic coordinate condition).
We thus arrive at (\ref{asv2}) and (\ref{asgam2}) but with 
remaining terms of the form $O^2(r^{-2-\epsilon})$. The required
falloff of $O^{2}(r^{-3})$ can be obtained by another
straightforward iteration. This ends the proof. \\

We remark that the iteration leading to (\ref{asv2}) and
(\ref{asgam2}) can in fact be continued to arbitrary high order
as in the Lorentzian case (\cite{SB}).

We also note that in terms of the radial coordinate $\widehat r$ employed
here the Schwarzschild metric takes on the asymptotic form
(\ref{Schwarz}), whereas the $\widehat r$ in the 
Taub-bolt metric in which the 
asymptotic form (\ref{asv2}), (\ref{asgam2}) holds 
is related to the $r$ used in (\ref{Taubbolt}) by
$\widehat r = r - \frac{5}{4}|n|$. 

We now introduce two scalars $\Omega_{\op}$ and $\Omega_{\om}$ 
and new metrics $g_{ij}^{\op}$ and $g_{ij}^{\om}$ on ${\cal N}^{\infty}$ as 
follows (recall that $0 < \Theta < 1$ on $\N^{\infty}$).
\begin{equation}
\label{gpm}
g^{\oo}_{ij} = \Omega_{\oo}^2 \gamma_{ij} =
\left ( \frac{1 \pm \sqrt{\Theta}}{2 \sqrt{\Theta}} 
\right )^2\gamma_{ij}  =
\frac{1}{16} \left [ \sqrt{ \left (1+ V^{2} \right)^2 -
\omega^2} \pm 2 V \right ]^2 g_{ij}.
\end{equation}
(Super- and subscripts ${\op}$,${\om}$ and ${\oo}$ on $g_{ij}$ and
$\Omega$ have nothing to do with the suffixes $+$, $-$ and $\pm$ on
$v$, $m$ and $m_{i}$ used before).

\begin{lemma}
\label{comp}
Let $({\cal M},g_{\mu\nu})$ satisfy conditions {\bf T1} and {\bf
T4} of the theorem. 

Then $({\cal N}, g_{ij}^{\op})$ is asymptotically flat with
vanishing mass. 

Assume further that the constants $m_{+}$ and $m_{-}$ of Lemma
\ref{as} do not vanish.

Then $({\cal N}, g_{ij}^{\om})$ has a compactification such that
$\widetilde {\cal N} = {\cal N} \cup \Lambda$ where $\Lambda$ is a point,
and $g_{ij}^{\om}$ has a $C^2-$ extension to $\Lambda$. 

\end{lemma}

{\it Proof.} 
Due to Lemma \ref{as}, the asymptotic behaviour
of $\Omega_{\op}$ and $\Omega_{\om}$ is
\begin{eqnarray}
\label{asOm}
\Omega_{\op}  =  1 + \frac{m^2 - n^2}{4\widehat r^2} + 
O^{2} \left (\frac{1}{\widehat r^{3}} \right),
\quad
\Omega_{\om}   =  \frac{m^2 - n^2}{4\widehat r^2} + 
O^{2} \left (\frac{1}{\widehat r^{3}} \right),
\end{eqnarray}
where $m = \frac{1}{2}(m_{+} + m_{-})$ and $n = \frac{1}{2}(m_{+} - m_{-})$.
The proof of the first part is trivial (see e.g. \cite{SY} for the
definition of mass of an {\it AF} manifold). 
As to the second part, if $m_{\pm} \ne 0$ then
in coordinates $\bar x^{i} = \widehat r^{-2}\widehat x^{i}$ it
is easy to see (as in the Lorentzian case, \cite{SB}) that $g^{\om}_{ij}$
has a $C^2-$ extension to the point ``at infinity''
$\Lambda$ given by $\bar x^{i} = 0$. \\

We remark that, again as in the Lorentzian case \cite{BSK}, there is even an
analytic compactification.

\section{The theorem}

Recall that $\omega_{\mu}$ is curl-free (\ref{Dom})
and hence there exists, locally on $\M$, a scalar $\omega$
defined by $\nabla_{\mu}\omega = \omega_{\mu}$.
Using assumption {\bf T3} of the theorem, $\omega$ exists globally on $\M$
(and hence also globally on $\N$) and is defined up to an additive
constant. Since $\N$ has a  single end, we choose
$\omega$ such that it vanishes at infinity. (It then coincides with the
scalar $\omega$ defined only on $\N^{\infty}$ in Sect. 3).

We can now introduce the ``Ernst potentials'' 
${\cal E}_{+} = V^{2} + \omega$ and 
${\cal E}_{-} = V^{2} - \omega$
which have the following global properties.
\begin{lemma}
\label{inequ}
Let $({\cal M}, g_{\mu\nu})$ satisfy conditions {\bf T1}, 
{\bf T3} and {\bf T4} of the theorem.

Then $-1 < {\cal E}_{\pm} \le 1$. More specifically, either
\begin{description}
\item[L1] at least one of the potentials satisfies ${\cal E}_{\pm} = 1$
on $\N$ or
\item[L2] both potentials satisfy $-1 <  {\cal E}_{\pm}  < 1$
on $\N$.
\end{description}

\end{lemma}

{\it Proof.}
Ricci flatness implies the following equations for
${\cal E}_{\pm}$ on $\N$ and on $\M$, respectively,
\begin{eqnarray}
\label{DDe}
D^{2} {\cal E}_{\pm} & = & V^{-2} 
D_{i}{\cal E}_{\pm} D^{i}{\cal E}_{\pm} \ge 0. \\
\label{dde}
\nabla^{2} {\cal E}_{\pm} & = & V^{-2} 
\nabla_{\mu}{\cal E}_{\pm}\nabla^{\mu}{\cal E}_{\pm} \ge 0.
\end{eqnarray}

The strong maximum principle (Theorem 3.5 in \cite{GT})
applied to (\ref{DDe}) shows that ${\cal E}_{\pm}$ can only have a maximum
inside $\N$ if it is constant.
Furthermore, since the fixed points of $\mu_{\tau}$  are of course interior
points of $\M$, the maximum principle applied to (\ref{dde}) in a
neighbourhood of the fixed points
excludes maxima at such points (on $\M$ and hence also on $\widehat \N$).
Since both potentials approach the value $1$ at infinity, we have 
either ${\cal E}_{\pm} < 1$ or ${\cal E}_{\pm} = 1$ on ${\cal N}$. 
Using this and the identity
${\cal E}_{\pm} = 2 V^{2} - {\cal E}_{\mp} > - {\cal E}_{\mp} \geq -1$
the remaining statement of the lemma follows easily.\\

Note that by imposing {\bf T4} we require $\widehat \N$ to be the
union of a compact set and an ``end''. We do not yet know whether this
property also holds for $\M$. Therefore, the maximum principle applied 
only to (\ref{dde}) globally on $\M$ would not suffice to prove the lemma.

We are now in the position of proving the theorem stated in the introduction,
which we copy here for convenience. \\ \\
{\bf Theorem}~{\it The Riemannian Schwarzschild and the
Taub-bolt instanton solutions are 
the only spaces $({\cal M}, g_{\mu\nu})$ such that 
\begin{description}
\item[T1.] ${\cal M}$ is a 4-dimensional manifold
with a Riemannian, Ricci-flat $C^{2}-$metric $g_{\mu\nu}$ 
which admits (at least) a 1-parameter group $\mu_{\tau}$ of
isometries. Moreover, the set $\N$ of non-trivial orbits of $\mu_{\tau}$
is a manifold.
\item[T2.] $\mu_{\tau}$ has no isolated fixed points on $\M$.
\item[T3.] $\M$ is simply connected.
\item[T4.] $(\N,g_{ij})$ is asymptotically flat (AF) and the
norm of the Killing field corresponding
to $\mu_{\tau}$ tends to a constant at infinity.
\end{description} } 
{\it Proof.}~The proof has two parts, {\bf 1} and {\bf 2}, which
correspond to the two cases of Lemma \ref{inequ}.\\ \\
{\bf 1.} {\bf L1}  of Lemma \ref{inequ} holds.\\

We can assume without loss of generality that
${\cal E}_{-}=1$ on ${\cal N}$, which implies $v_{-} = 0$
and $v_{+} = V^{-2} -1$, and hence  $A_{i}=0$
and $\Theta=1$. This means that equations (\ref{Lapv}) 
and (\ref{gamRicc}) become  $D_{i} D^{i} V^{-2} = 0$
and $R_{ij} = 0$, i.e. $\gamma_{ij}$ is locally flat.  Using
coordinates adapted to the Killing vector, the metric takes the 
local form
\begin{equation}
ds^2 = V^{2} \left (d\tau + \eta_{i} dx^i \right )^2 +
V^{-2} \delta_{ij} dx^i dx^j.
\label{locform}
\end{equation}
The one-form  $\bm{\eta}=\eta_{i}dx^i$ on $\N$ can be obtained 
from $\omega$ as any particular solution of 
$\mbox{curl}_{\gamma} \, \bm{\eta} =- V^{-4} \mbox{grad} \, \omega$.
Since ${\cal E}_{-}=1$ we have $\omega = V^{2} -1$ and therefore
$\mbox{curl}_{\delta} \, \bm{\eta} =  \mbox{grad} \, V^{-2}$.
Let us now show that $(\M,g_{\mu\nu})$ must contain nuts, against
our assumptions, or must be the four-dimensional Euclidean space.

Consider first the case when ${\cal L}_{\M}$ is empty. Then it follows from 
the definition of asymptotic flatness that $\N = \widehat \N$ is a
complete Riemannian manifold. Since, moreover,
$\gamma_{ij}$ is flat on the asymptotically flat end $\N^{\infty}$,
$(\N, \gamma_{ij})$ must be diffeomorphic to $(R^{3},\delta_{ij})$.
Furthermore, $V^{-2}$ is well-defined everywhere on $\N$ and solves
the flat Laplace equation $\triangle V^{-2}  = 0$. Hence
$V$ must be constant and equal to its asymptotic value 1.
This implies $\eta_i dx^i = 0$ and therefore $(\M,g_{\mu\nu})$ is locally flat.
But since $\M$ is simply connected, the flatness of
$(\M,g_{\mu\nu})$ follows. 

We have thus shown that $(\M,g_{\mu\nu})$ must contain either nuts or bolts,
or else the space is Euclidean. Assume that $(\M,g_{\mu\nu})$
contains a bolt ${\cal B}$ with gravity $\kappa$. It is easy to see that 
$\nabla_{\mu} \xi_{\nu} \nabla^{\mu} \xi^{\nu}|_{\cal B} =
2\kappa^2$ and $\epsilon_{\mu\nu\alpha\beta} \nabla^{\mu} \xi^{\nu}
\nabla^{\alpha} \xi^{\beta}|_{\cal B} =0$. Using (\ref{gRicc}),
we obtain that the Ricci tensor ${\cal R}$ of the metric 
$g_{ij}=V^{-2}\delta_{ij}$ is
\begin{eqnarray}
{\cal R} = \frac{3}{8 V^4}
\nabla_{\mu} {\cal E}_{+}
\nabla^{\mu} {\cal E}_{+} = \frac{3}{8 V^{2}} \left (
2 \nabla_{\mu} \xi_{\nu} \nabla^{\mu} \xi^{\nu} +
\epsilon_{\mu\nu\alpha\beta} \nabla^{\mu} \xi^{\nu}
\nabla^{\alpha} \xi^{\beta} \right ),
\label{sing}
\end{eqnarray}
where the first equality requires  $\omega = V^{2} -1$
and the second equality is generally valid.
Hence, ${\cal R}$ must be singular
on the bolt, which is impossible from Lemma \ref{bound}.
We can therefore conclude that {\bf L1} in
Lemma \ref{inequ} requires that
the four-metric contains nuts or else $(\M,g_{\mu\nu})$ is flat.
\\ \\
{\bf 2.} {\bf L2} in lemma \ref{inequ} holds.\\

We first observe, using Lemma \ref{inequ}, that the fields 
$v_{\pm}=(1 + {\cal E}_{\pm})^{-1}(1 - {\cal E}_{\pm})$
introduced on $\N^{\infty}$ in Sect. 3 are well defined and
non-negative on all of $\N$, and the same applies to
 $\Theta = 1 - v_{+}v_{-} =  4 V^{2} \left (1+ {\cal E}_{\pm} \right)^{-1}
\left (1+ {\cal E}_{\mp} \right )^{-1}$.

We also note that the constants $m_{\pm}$ in Lemma \ref{as}
cannot vanish. Assume, for the contrary, that e.g. $m_{-} = 0$.
Assume also for the moment that $m_{i}^{-}$ in (\ref{asv2}) does not vanish.
The term of order $\hat{r}^{-2}$ in (\ref{asv2}) containing
this constant does not
have a definite sign and dominates the expansion for large $\hat{r}$,
which contradicts $v_{-} > 0$. Therefore, $m_{i}^{-} = 0$. 
Applying Lemma \ref{Lapinv} to (\ref{Lapv}) we find that the leading
term in the expansion of $v_{-}$ must necessarily be a solution
of the flat Laplace equation. Since all such solutions of order
$\hat{r}^{-p}$, $p \ge 2$, change sign on $\N^{\infty}$,
the same argument as above leads to a contradiction unless 
$m_{-} \ne 0$, and in the same way we conclude that $m_{+} \ne 0$. 
Hence Lemma \ref{comp} on the asymptotic structure applies.

We next show that $g_{ij}^{\oo}$ are regular on 
$\partial \N = {\cal L_{\NN}}$. 
Since $V \Omega_{\op}|_{\partial \N} =
V \Omega_{\om}|_{\partial \N}$
the metrics ${}^{2}g_{ij}^{\oo} = g_{ij}^{\oo} - n_{i}^{\oo} n_{j}^{\oo}$ 
agree on the bolts $\partial \N$ (the unit normal vectors of $\partial \N$  
with respect to $g_{ij}^{\oo}$ are denoted by $n_{i}^{\oo}$).
We also notice that under a conformal rescaling $g_{ij}' =
\Omega(V,\omega)^{2}~g_{ij}$, the extrinsic curvature
$k_{ij}$ of a 2-dimensional
submanifold ${\cal S}$ in $({\cal N}, g_{ij})$ changes according to
\begin{equation}
\label{extr}
k_{ij}' = \Omega k_{ij} + 
\Omega^{-2} ({}^{2}g_{ij}') n^{k} 
\left( \frac{d\Omega}{dV} {\cal D}_{k} V +
       \frac{d\Omega}{d\omega} {\cal D}_{k} \omega \right ),
\end{equation}
where $n^{k}$ is the unit outward normal of ${\cal S}$ with
respect to $g_{ij}$.

Setting ${\cal S} = \partial \N$ we have $k_{ij} = 0$ due to
lemma \ref{bound}. We next insert $\Omega = V \Omega_{\oo}$ in 
(\ref{extr}) and note that (\ref{gLapom}) can be written as 
$\nabla^{2} \omega = 4 V^{-1} \nabla_{\mu} V \nabla^{\mu} \omega$
on ${\cal M}$, and so 
$g^{ij}{\cal D}_{i}V {\cal D}_{j}\omega|_{\partial \cal N} = 0$.
Hence the second fundamental forms $k_{ij}^{\pm}$ of $\partial \N$ in 
$(\N, g^{\oo}_{ij})$ satisfy $k^{\op}_{ij} = - k^{\om}_{ij}$. 
Therefore we can glue together the two 
manifolds $(\N,g^{\op}_{ij})$ and $(\widetilde \N, g^{\om}_{ij})$ 
along $\partial \N$ to obtain a $C^2$ manifold with $C^1$ metric 
\cite{CD}. 
Since the metric is piecewise $C^2$, it follows that it is
$C^{1,1}$. 
By Lemma \ref{comp}, the resulting space is a complete three-dimensional 
asymptotically flat manifold with $C^{1,1}$ metric and vanishing mass. 
A short computation shows that it also has non-negative Ricci scalar, namely
\begin{equation}
\label{Rpm}
R^{\oo} \Omega^4_{\oo} = 
\frac{3 \pm 2 \sqrt{\Theta}}{2 \Theta^3} \gamma^{ij}A_i A_j \geq 0.
\end{equation}

 The rigidity part of the positive mass theorem \cite{SY} implies that 
this manifold must be diffeomorphic to $R^3$ with the flat metric. In 
particular, both metrics $g^{\oo}_{ij}$ are flat. 

Expression  (\ref{Rpm}) also shows that $A_{i}=0$, which
is equivalent to $v_{+} = \alpha v_{-}$, where $\alpha$
is a positive constant (due to $v_{\pm} > 0$). It is convenient to
introduce a function $H = (1 + \sqrt{\Theta})/(1 - \sqrt{\Theta})$
which is regular and satisfies $H|_{\N} > 1$ and $H |_{\partial
\N} = 1$. Eqn. (\ref{gamRicc}) provides the Ricci tensor
of the metric $\gamma_{ij}$ in terms of the gradient of $v_{+}$.
Using (\ref{gpm}) and the fact that $g_{ij}^{(+)}$ is flat,
the standard formula for the Ricci tensors of conformally related
metrics gives (written in Euclidean coordinates):
\begin{eqnarray}
\partial_i \partial_j H - 
\frac{\delta^{kl} (\partial_k H)(\partial_l H)}
{2H} \delta_{ij}= 0,
\label{DDH}
\end{eqnarray}
Multiplying (\ref{DDH}) with $\partial_{j}H$ we obtain $\delta^{lk} \partial_l H
\partial_k H = \beta H $, where $\beta $ is a constant. Since 
$H$ cannot be constant it follows that $\beta > 0$, and we
can write $\beta = 16 M^{-2}$ for some positive constant $M$.
Then (\ref{DDH}) becomes $\partial_i \partial_j H = 8 M^{-2} \delta_{ij}$
whose general solution can be written,
after performing an appropriate translation $\tilde x^i = x^i + c^i$,
as $H = 4 M^{-2} \delta_{ij} \tilde x^i \tilde x^j$. 
The knowledge of $H$ implies that
of $v_{+}$ and $\Omega_{(+)}$. In spherical coordinates 
$\{\tilde{r}, \tilde{\theta}, \tilde{\phi} \}$, we have
\begin{eqnarray*}
\label{vgam}
v_{+} = \sqrt{\alpha} \frac{4 M \tilde{r}}{M^2 + 4 \tilde{r}^2},
\hspace{4mm}
ds^2_{\gamma} |_{\N} = \left (1 - \frac{M^2}{4 \tilde{r}^2} \right )^2
\left ( d\tilde{r}^2 + \tilde{r}^2 d\Omega^2 \right ),
\hspace{4mm} \tilde{r} > M/2,
\end{eqnarray*}
where we used $H|_{\partial\N}=1$ and $H|_{\N}>1$ to determine the
range of $\tilde{r}$. We now define two constants 
$m > |n| \geq 0$ by $\alpha= (m + n)^{-1} 
(m - n)$ and $M = \sqrt{\alpha} (m + n)$ and perform the coordinate
transformation $r = m + \tilde{r} + M^2/(4 \tilde{r}^2)$.  
The metric $\gamma_{ij}$, $V$ and $\omega$ take the form
\begin{eqnarray}
\label{gam}
ds_{\gamma} & = & dr^{2} + \left( r^{2} - 2mr + n^{2}
\right) d\Omega^{2} \qquad r > m + \sqrt{m^{2} - n^{2}}\\
\label{Vom}
V^{2} & = & \frac{r^2 - 2 m r + n^2}{r^2 - n^2} \qquad
\omega = \frac{2 n (r - m)}{(r^2 - n^2)}
\end{eqnarray}

Using this, the metric $g_{\mu\nu}$ can be reconstructed by solving 
$\mbox{curl}_{\gamma} \, \bm{\eta} =- V^{-4} \mbox{grad} \, \omega$
as in part {\bf 1}. Finally, global regularity of $g_{\mu\nu}$ 
on ${\cal M}$ requires \cite{DP} that either $n = 0$ which gives the 
1-parameter family of Schwarzschild instantons, or that $m = \frac{5}{4}|n|$ 
which gives the 1-parameter family of Taub-bolt instantons.

\section{Discussion}
We can think of improving our theorem in various directions.
We discuss here briefly three of the most striking problems, namely:
\begin{description}
\item[P1.] As mentioned already in the introduction, we believe
that in {\bf T1} the assumption that $\N$ is a manifold can be dropped.
\item[P2.] We would like to prove uniqueness of Taub-bolt purely
in the 4-dimensional setting, e.g. as formulated in the
conjecture in the introduction.
\item[P3.] It would be desirable to have a uniqueness result for the case
in which nuts as well as bolts are a priori allowed to be
present. 
\end{description}

{\bf P1.} We can show that $\N$, the space of non-trivial Killing orbits
$\mu_{\tau}$ of $\M$  is in fact a manifold provided that all 
these orbits have the same period. In a neighbourhood of a fixed
point $p$ of $\mu_{\tau}$, the
period of the orbits is determined by the ``gravities'' of
the fixed point (c.f. \cite{GH}). This is due to
 the commutativity of the isometry and the
exponential map, viz. $\exp (\mu_{\tau \ast} X) =
\mu_{\tau}(\exp X)$ for a tangent vector $X$ at $p$, and to the
fact that locally the exponential map is one-to-one.
Since a bolt has only one ``gravity'', all orbits in its
neighbourhood have the same period. More generally, in a
spacetime with a bolt one has control over the period of the
orbits on a domain in which the exponential map is non-singular.
This domain is bounded by the cut points of the geodesics emanating
from the bolt (c.f. Sect. VIII, Theorem 7.4. of \cite{KN2}).
On these cut loci, the period might change and the manifold
structure of $\N$ may be lost, though we do not believe that
this will happen.\\

{\bf P2.} 
In the {\it AF} case, the conjecture can actually be proven rather
straightforwardly along the lines of the Lorentzian case
\cite{MH} (and leads to the Schwarzschild instanton
(\ref{Schwarz})), by first showing an analog of the ``staticity
theorem'' (c.f. Sect. 8.2 of \cite{MH}), followed by the reasoning
of the present paper restricted to the case of a
hypersurface-orthogonal Killing vector. 

As to the asymptotically locally flat case, we
recall here from \cite{GPRP} the definition of {\it ALF}
(with slight modifications, in particular using the
cyclic group $Z_{Q}$ instead of the more general options 
in \cite{GPRP}).

\begin{enumerate}
\item
The ``end'' $\M^{\infty}~\setminus$ \{a suitably chosen compact set\}
is diffeomorphic to $R \times \left (S^{3}/Z_{Q}\right)$
\item 
The lift $d\check s^{2}$ of the line element $ds^{2}$ to
the covering space $R \times S^{3}$ takes the form
\begin{equation}
\label{asg} 
d\check s^{2} = dr^{2} + r^{2}\left(\sigma_{1}^{2} + \sigma_{2}^{2}\right) +
\sigma_{3}^{2} + O^{2}(r^{-1}),
\end{equation}
where $r \in (r_{0}, \infty)$ for some constant $r_{0}$ is a
radial coordinate (in the $R$ direction), $\sigma_{i}$ ($i
= 1,2,3$) are the left-invariant one-forms 
on the unit sphere.
\end{enumerate}

If a 1-parameter group of isometries
$\mu_{\tau}$ of $({\cal M}^{\infty}, g_{\mu\nu})$ is present, it
is likely that the latter will leave the lens spaces 
$S_{r}/Z_{Q}$ invariant (with  $S_r$ denoting the
$3$-spheres of constant radius $r \in (r_{0}, \infty)$), and that
$Z_{Q}$ will act on each of the $S_r$ as a
subgroup of the lift $\check \mu_{\tau}$ of $\mu_{\tau}$ to the $S_r$.

If this is the case, we can show that the 1-form dual to the
Killing field $\xi^{\mu}$ is parallel to $\sigma_{3}$, and that the quotient
manifold $(\N, g_{ij})$ is asymptotically flat, as defined in
Sect. 3 and as required in our theorem.

We remark that the ``end'' of the Taub-bolt metric is 
$R \times S^{3}$ . So we could simplify the asymptotic condition
by requiring this topology instead of the lens space  $S_r/Z_Q$.\\

{\bf P3.} 
The theorem given above only allows the presence of
bolts. Assume we relax assumption {\bf T2} by
also allowing (or only allowing) 
the presence of nuts, but we keep the other requirements.
Then there is the example of multi-Taub-nut space
(which has only nuts) and which might well be the only example.
Since this space has a conformally flat space of orbits,
it might then be possible to show its uniqueness by the strategy
of Bunting and Masood-ul-Alam \cite{BM} as pursued in the present paper.
If we allow bolts a priori, the problem with this strategy is to
find a suitable conformal rescaling of the metric on $\N$ which
in particular keeps the metric regular on the bolts. In general
a rescaling might shrink the bolts to points or shift them to
infinity. There is nevertheless hope to
get sufficient control also in this general case in order to
apply a positive mass theorem.

On the other hand, the uniqueness result for the Euclidean Kerr metric
\cite{WS} is restricted to the asymptotically flat case,
allows only nuts and effectively restricts their
number to 2 by requiring that ${\cal M} =  R^2 \times S^2$.
We note that in this case the space of orbits is not a manifold
(contrary to the statement in \cite{WS}. The proof of \cite{WS}
needs to be and can easily be rewritten accordingly).
The strategy of this proof is based on a generalization of
Israel's proof of uniqueness of the Schwarzschild solution,
and on a suitable characterization of the Kerr metric.
In fact, this characterization naturally extends to the
asymptotically locally flat class of ``Kerr-Taub-NUT'' instantons
\cite{WS} (called ``Kerr-Taub-bolt'' metrics by the discoverers
\cite{GPLP} since they refer to another Killing vector).
Assuming {\it ALF} instead of {\it AF} it should be possible to extend the 
uniqueness proof of \cite{WS} to this class. While we also believe that the
topological conditions of the original proof can be relaxed,
they can probably not be removed altogether.

It would also be interesting to prove uniqueness of asymptotically (locally)
Euclidean instantons \cite{GPRP} along these lines.\\ \\
{\bf Acknowledgement} We are grateful to Piotr Chru\'sciel for extensive
comments on a first draft of this work, which lead to substantial corrections
and improvements, 
and Robert Beig and Helmuth Urbantke for helpful discussions.

\end{document}